# AttendAffectNet: Self-Attention based Networks for Predicting Affective Responses from Movies

Ha Thi Phuong Thao*, Balamurali B.T.*, Dorien Herremans*, Gemma Roig†
*Singapore University of Technology and Design, 8 Somapah Rd, Singapore 48737
†Goethe University Frankfurt, 60323 Frankfurt, Germany
Email: thiphuongthao_ha@mymail.sutd.edu.sg, {balamurali_bt, dorien_herremans}@sutd.edu.sg, roig@cs.uni-frankfurt.de

*Abstract*—In this work, we propose different variants of the self-attention based network for emotion prediction from movies, which we call AttendAffectNet. We take both audio and video into account and incorporate the relation among multiple modalities by applying self-attention mechanism in a novel manner into the extracted features for emotion prediction. We compare it to the typically temporal integration of the self-attention based model, which in our case, allows to capture the relation of temporal representations of the movie while considering the sequential dependencies of emotion responses. We demonstrate the effectiveness of our proposed architectures on the extended COGNIMUSE dataset [1], [2] and the MediaEval 2016 Emotional Impact of Movies Task [3], which consist of movies with emotion annotations. Our results show that applying the self-attention mechanism on the different audio-visual features, rather than in the time domain, is more effective for emotion prediction. Our approach is also proven to outperform many state-of-the-art models for emotion prediction. The code to reproduce our results with the models' implementation is available at: https://github.com/ivyha010/AttendAffectNet.

## I. Introduction

The emotional impact of movies has been long studied by psychologists [4], [5], [6]. Watching a movie is typically an emotional experience, whereby each scene or sound effect can evoke different types and levels of emotion [7]. Building a system to automatically predict emotions evoked in viewers from watching a particular movie would offer a useful tool to producers in the film and advertisement industry. Most studies focus on recognizing emotions of humans that appear in videos on the basis of their facial expressions and spoken audio [8], [9], [10] rather than predicting emotions that the viewers are experiencing based on the audio and video of movies.

Researchers in psychology have proposed many models to capture human emotions, which can broadly be divided into two major groups: categorical [11], [12] and dimensional [13], [14]. In many affective computing studies [1], [15], [16], [17], human affective responses are typically represented in the arousal and valence dimensions of the circumplex model of affect proposed by Russell [18]. Arousal represents the degree of energy, while valence describes the degree of pleasantness. In addition to arousal and valence, one more dimension of the circumplex model of affect namely dominance [18], which refers to the sense of "control" or "attention", is sometimes also used to represent affective responses of viewers [19]. Our aim is to use a quantitative model to predict valence and arousal values using only the movies as inputs.

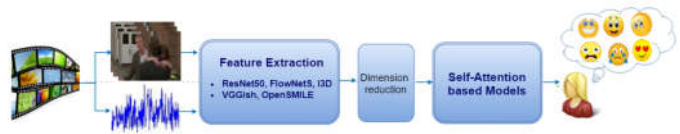

Fig. 1. Overview of our approach for predicting emotions from movies.

There are current studies [20], [21], [22], [23], [24] that have demonstrated promising results in affective response prediction from movies. However, most of them mainly focus on using fusion techniques such as early fusion, late fusion, etc. and they do not explicitly tackle the relation among multiple modalities, which change over time. In order to tackle these issues, we introduce a model with the self-attention mechanism from the Transformer model [25] to predict emotion, which we call AttendAffectNet. The self-attention mechanism can capture temporal dependencies [25] as well as spatial dependencies [26] of an input sequence. Therefore, we apply it to predict affective responses of viewers, which are represented in terms of valence and arousal (as defined in Russel's circumplex model of affect). We explore two variants of the self-attention based model. The first one, named Feature AttendAffectNet, applies the self-attention mechanism on features extracted from audio and video streams of movies. The second model, named Temporal AttendAffectNet, uses the self-attention mechanism on the time domain of the movie taking all the concatenated features as an input at each time step.

An overview of our proposed approach is illustrated in Figure 1. Our AttendAffectNet is used on input video and audio features, extracted with pre-trained convolutional neural network architectures (CNN) including the ResNet-50 network [27] and RGB-stream I3D network [28] for appearance feature extraction, FlowNet Simple (FlowNetS) [29] for motion features, VGGish neural network [30] and the OpenSMILE toolkit [31] to extract audio features.

We evaluated our proposed AttendAffectNet models on the extended COGNIMUSE dataset [1], [2] and MediaEval 2016 Emotional Impact of Movies Task [3]. Our experimental results reveal the importance of audio features in driving viewer affect. Also, applying the self-attention mechanism on the features for emotion prediction obtains a higher accuracy than using it on time domain.

## II. RELATED WORK

Predicting affective responses of viewers evoked by videos using only the videos themselves has a large range of applications, such as in film making and advertising industries. Yet, most of the existing literature is on predicting the emotional expressions of humans appearing in videos [32], [33]. Studies in emotion recognition of humans in videos have applied a multimodal approach using audio, visual and text features [34], [35], [36].

Recent methods to predict evoked emotion directly from movies also use a multimodal approach [20], [23], [24]. Deep convolutional neural networks (CNNs) such as VGG [37], ResNet-50 network [27] and Inception-v3-RGB [38], [39] have been deployed to obtain semantic features from image frames for predicting emotions of viewers evoked by movies [24], [40], [41].

In addition to appearance features of objects extracted from still frames, features related to motion offer pivotal cues for emotion prediction. In [24], ResNet-101 network [27] pre-trained on the ImageNet dataset [42] with its first convolutional layer and last classification layer fine-tuned on UCF-101 dataset [43] is used to extract motion features from optical flows, which are estimated using PWC-Net [44] pre-trained on MPI Sintel final pass dataset [45]. In [41], authors use Inception-v3-Flow [38], [39] to obtain motion features.

In our study, we make use of the pre-trained ResNet-50 network to extract appearance features from still RGB frames. To obtain appearance features from a stack of consecutive frames while taking the temporal relation cross multiple frames into account, we use the pre-trained I3D model, which achieves state-of-the-art performance on a large number of action recognition benchmarks [28]. We also extract optical flow features, as the motion information is important for predicting the viewer affect, as shown in [24], [46]. We use the FlowNet Simple (FlowNetS) network pre-trained on the Flying Chairs dataset [29].

Importantly, many studies have shown that there exists a high correlation between speech/music and emotion [47], [48]. Audio features can be extracted either using toolkits such as YAAFE [49], OpenSMILE [31] or by making use of network architectures based on spectrograms or waveforms such as AlexNet, VGGish, Inception and ResNet [30], as well as SoundNet [50]. OpenSMILE-extracted features have shown to be particularly effective for emotion prediction models [24], [51]. VGGish is also used in [41] for audio feature extraction. In [52] audio features extracted using VGGish outperform Mel-frequency cepstral coefficients (MFCCs), SoundNet and OpenSMILE-extracted features in the INTERSPEECH 2010 Paralinguistic challenge [53]. We therefore opt to include both VGGish and OpenSMILE-extracted features in our study.

To build the multimodal model to predict emotions from videos, a wide range of unsupervised and supervised approaches have been developed. Hidden Markov models (HMMs) are proposed in the extended version of the COG-NIMUSE dataset [1], [2] to predict arousal and valence separately at the frame level.

In [20], movies in the extended COGNIMUSE dataset are split into non-overlapping 5-second segments, in which the arousal and valence values are averaged over all frames in each segment. The authors predict valence and arousal for every 5-second segment instead of every frame using simple linear regression with fusion techniques. A follow-up approach conducted in [23] uses LSTM accompanied with correlation-based feature selection [54] and outperforms the approach in [20]. The LSTM is also used to develop an adaptive fusion recurrent network in [41] in the MediaEval 2016 Emotional Impact of Movies Task [3]. This network outperforms SVR in [22], [55] and Random Forest in [55], LSTM and Bidirectional-LSTM in [21], Arousal-Valence Discriminant Preserving Embedding algorithm [56].

In this work, we propose to use the Transformer model for evoked emotion prediction from movies, which is able to capture both the correlation among multiple modalities as well as temporal inputs. It also has been shown to outperform LSTM in many different tasks, such as text translation tasks [57], speech emotion recognition [58]. This model is also applied in human action recognition task and has a good performance [59].

Inspired by these findings, in this paper, we explore self-attention based models for predicting emotions of viewers from movies. We report results on COGNIMUSE dataset and MediaEval 2016, which are the most commonly used benchmarks for this task.

## III. ATTENDAFFECTNET ARCHITECTURE

We propose the AttendAffectNet model, which is based on the self-attention mechanism, to predict emotions in terms of valence and arousal. The network uses features extracted from both video and audio streams. In this section, we first explain the model inputs, i.e. video and audio features, then describe the model.

### A. Video features

We extract video features at the frame level. Therefore, we first use the FFMPEG tool [1] to obtain $T$ frames from each video. For doing so, we extract one frame for every $\frac{t_i}{T}$ seconds, where $t_i$ is the duration of the $i$-th clip, following a similar frame extraction procedure as in [46], [59].

*a) Appearance features:* We first extract the static appearance features of objects from still frames using the ResNet-50 network [27] with pre-trained weights on the ImageNet dataset [42]. For every image passed through all layers of the ResNet-50 network, except for the last fully connected layer, we obtain a 2048-feature vector. An element-wise averaging of the extracted features over all frames is computed to finally obtain a 2048-feature vector for each movie excerpt.

We also use the RGB-stream I3D model [28] based on Inception-v1 [60] with pre-trained weights on the Kinetics human action video dataset [61] to extract spatio-temporal features from video frames carrying information about the appearance and temporal relation. The stack of $T$ frames corresponding to an input size of $C \times T \times H \times W$ (where $C$ is the number of channels, $H$ and $W$ are the frame height and width respectively) is passed through the pre-trained Inception-v1 based RGB-stream I3D network after removing all layers following its "mixed-5c" layer. As a result, we obtain a feature map of size $T' \times H' \times W' = 1024 \times \frac{T}{8} \times \frac{H}{32} \times \frac{W}{32}$. We then apply the average pooling with the kernel size of $\frac{T'}{8} \times \frac{H}{32} \times \frac{W}{32}$ to get a 1024-feature vector from each movie excerpt.

[1] https://www.ffmpeg.org/

*b) Motion features:* The cost of optical flow estimation is relatively expensive. Therefore, instead of extracting motion features from optical flows, in this study we aim to make use of a network that predicts optical flow to obtain low-level motion information. The FlowNetS consists of contracting and expanding parts [29], which make it have an encoder-decoder-like structure. We use its contracting part as a feature extractor to obtain motion features, in which the FlowNetS was pre-trained on the Flying Chairs dataset [29]. A 1024-feature vector is obtained by passing each pair of consecutive frames through the encoder of the FlowNetS. We compute the element-wise average of the extracted motion features over all pairs of frames to get a 1024-feature vector for each movie excerpt.

### B. Audio features

The OpenSMILE-extracted features have proven to provide meaningful audio features for emotion prediction tasks [24], [52]. We extract audio features using the OpenSMILE toolkit [31]. Additionally, we use the pre-trained VGGish neural network [30].

*a) OpenSMILE features:* The configuration file "emobase2010" in the INTERSPEECH 2010 paralinguistics challenge [53] is used to extract $1,582$ features from each 320-ms audio frame with a hop size of 40 ms. The extracted feature set consists of low-level descriptors including jitter, loudness, pitch, MFCCs, Mel filter bank, line spectral pairs with their delta coefficients, functionals, the duration in seconds, and the number of pitch onsets [62]. We compute the element-wise averaging of the OpenSMILE features over all 320-ms frames to get a $1,582$-feature vector for each movie excerpt.

*b) VGGish model:* We use the VGGish network, pre-trained for sound classification on the AudioSet dataset [63], as an audio feature extractor. From each 0.96-second audio segment, we obtain 128 audio features. We compute the element-wise averaging of the extracted features over all segments to finally obtain a 128-feature vector for each movie excerpt.

### C. Self-attention based models for emotion prediction

We first briefly review the Transformer architecture, and then describe our proposed self-attention based models.

*a) Transformer network:* The Transformer model relies on the self-attention mechanism, which relates different positions of a sequence in order to compute a representation of the sequence [25]. Instead of using a single attention function, the Transformer uses multi-head attention, in which the input is mapped into queries (Q), keys (K) and values (V) of dimensions $d_q, d_k, d_v$ respectively (where $d_q = d_k$) using different linear projections multiple times (denoted as $h$ times, which are also referred as $h$ heads). Note that the linear projections are fully connected layers, which are learned in the network training process. The attention function is then performed on queries, keys and values in parallel so as to create output values of $d_v$ dimensions each.

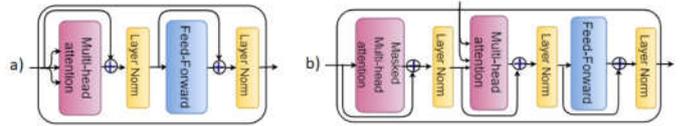

Fig. 2. a) Encoder and b) Decoder of the Transformer network with self-attention mechanism.

There are several attention mechanisms [64], [65], [66], among which additive attention [65] and dot-product attention [66] are the most two popular ones. The Transformer uses the scaled dot-product attention, which is the dot-product attention with a scaling factor $\frac{1}{\sqrt{d_k}}$ as in the equation below:

$$\text{Attention}(Q, K, V) = \text{softmax}\left(\frac{QK^T}{\sqrt{d_k}}V\right). \quad (1)$$

The $d_v$-dimensional output values are then concatenated and linearly projected to obtain the final values. We refer the reader to [25] for more details.

The Transformer includes an encoder-decoder framework, much like many other sequence translation models [65], [67]. The encoder converts the original input sequence into a sequence of continuous representations, and the decoder then generates an output sequence of symbols, in which one element is generated at a time. At each step, the previously generated symbols are used as an additional input of the model to generate the next one.

Since the Transformer does not use recurrence or convolution, it takes the order of the sequence into account by adding positional encodings to embedding inputs and previous outputs. The encoding $\overrightarrow{PE}_{pos} \in \mathbf{R}^d$ corresponding to position $pos$ is conducted using sine and cosine functions of different frequencies [25] as the following:

$$\overrightarrow{PE}_{pos}^{(i)} = \begin{cases} \sin(\omega_k . pos) & \text{if } i = 2k \\ \cos(\omega_k . pos) & \text{if } i = 2k+1, \end{cases} \quad (2)$$

where $\omega_k = \frac{1}{10000^{2k/d}}$, and $d$ is the encoding dimension, $i = 0, \ldots, d$.

The encoder and decoder of the Transformer network are illustrated in Figure 2. The encoder includes a stack of identical layers. Each layer consists of two sub-layers: a multi-head self-attention and a feed-forward neural network. Each sub-layer is surrounded by a residual connection [27] followed by a layer normalization [68].

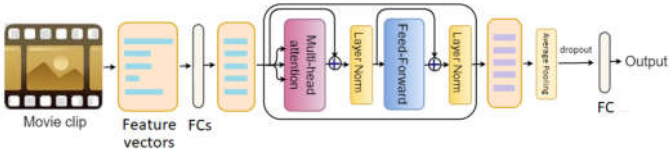

Fig. 3. Our proposed Feature AttendAffectNet.

The decoder also includes a stack of identical layers, however, each layer includes three sub-layers: masked multi-head self-attention, multi-headed attention and a feed-forward neural network. The masked multi-head self-attention sub-layer is designed to prevent positions from attending to subsequent positions by masking the future positions. The multi-head attention sub-layer in the decoder works like multi-head self-attention in the encoder, except its queries, $Q$, are created from the sub-layer below it, and its keys, $K$, and values, $V$, are from the output of the encoder. The feed-forward neural networks in both encoder and decoder consist of two linear transformations with a ReLU activation inserted in between. Each sub-layer is also surrounded by a residual connection followed by a layer normalization (we refer to [25] for more details).

Inspired by multi-head attention mechanism in the Transformer model, we apply the self-attention mechanism to capture the relation among features extracted from multiple modalities as well as temporal representations of the movie, which is crucial for predicting emotions of movie viewers. We propose the following variants of self-attention based models.

*b) Feature AttendAffectNet model:* Using different features is crucial for obtaining a high accuracy of emotion prediction, as has been shown in previous research [1], [36], [24]. Each feature has a distinctively important power in emotion prediction, as reported in the analysis by [24]. Inspired by these findings, we propose a model that uses the self-attention mechanism on the extracted audio-visual features carrying information about the appearance of objects, motion and sounds extracted from the whole movie clip. This also allows to incorporate the interactions among the features which might be also crucial for this task.

Each of five extracted feature vectors are passed through a fully connected layer of eight neurons for dimension reduction, before being fed to the encoder of the Transformer model. The Transformer encoder output consists of a sequence of five encoded feature vectors of eight elements each. We then apply average pooling over all of the encoded feature vectors to obtain an $8-$dimensional vector. A dropout is then applied on this vector followed by a fully connected layer to obtain the final output. The model is illustrated in Figure 3.

*c) Temporal AttendAffectNet model:* We explore another model using the self-attention mechanism on the time domain of movies while considering the sequential dependency of affective responses of movie viewers. This model consists of only the the decoder of the Transformer network.

To process the input, each original movie clip is first split into segments of the same length. For each temporal segment of the movie, we extract feature vectors from both audio and video streams. Each feature vectors is linearly mapped to the same dimension using a fully-connected layer of eight neurons. We then concatenate all of these dimension-reduced feature vectors to serve as the decoder input for one movie segment. In order to include information about the sequential order, we add a positional encoding vector to every concatenated dimension-reduced feature vector corresponding to each movie segment. Positional encoding is conducted by using the formula (2) with $d = 40$, which is the dimension of the concatenated dimension-reduced feature vector corresponding to each movie segment.

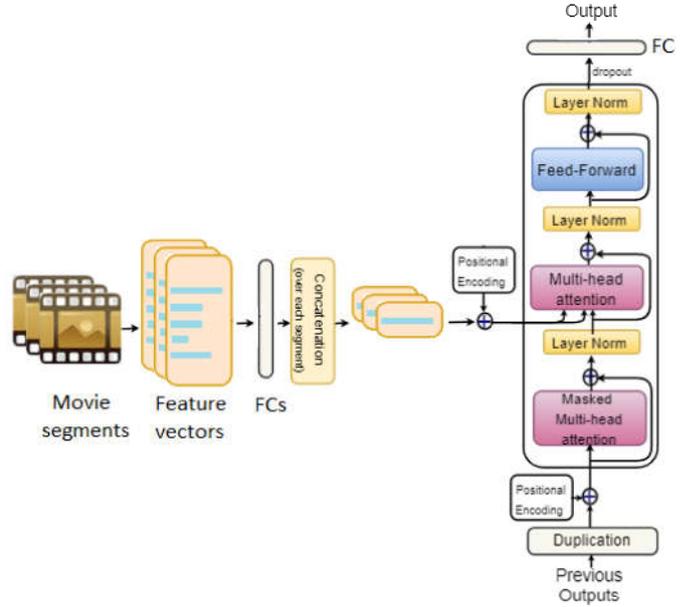

Fig. 4. Our proposed Temporal AttendAffectNet.

The decoder provides an output for each movie segment. The previous outputs are also used as an additional input to the decoder to predict the next one. Instead of using an embedding layer to embed the output word as described in the original Transformer model, we duplicate each previous scalar output (a single arousal/valence value) to obtain a vector of the same dimension as the concatenated dimension-reduced feature vector that represents each movie segment. The same positional encoding is also added to each previous duplicated output to predict the next one. A dropout is applied on the output vectors of the decoder before they are fed to a fully connected layer with only one neuron to obtain a sequence of final outputs. During the model training, however, we only back-propagate the error for the last output in this sequence, i.e. the sequence-to-sequence task is transferred to a sequence-to-one problem. This Temporal AttendAffectNet model is illustrated as in Figure 4.

*d) Feature and Temporal AttendAffectNet model:* We also propose a combination of the aforementioned two self-

attention based models, in which in the first stage we apply the self-attention mechanism on feature vectors extracted from each temporal segment of movie, and in a second stage, after the feature self-attention, we derive the idea from the Temporal AttendAffectNet to capture sequential dependency of emotions. However, this model does not perform better than the aforementioned ones. Hence, due to the space constraint, we refer the reader to the Appendix for more details on this model, as well as the results obtained.

## IV. EXPERIMENTAL SET-UP

We have conducted experiments to evaluate the performance of models described above on the task of predicting emotions of viewers from movies. The datasets and experimental implementation are described below.

### A. Datasets

We report results on the extended version of the COGNIMUSE dataset [1], [2] and the MediaEval 2016 Emotional Impact of Movies Task [3].

*a) Extended COGNIMUSE dataset:* This dataset [1] includes twelve continuous Hollywood movie clips of half an hour each with a frame rate of 25 fps, in which seven of them are from the COGNIMUSE dataset [2]. Emotion is annotated at frame level, in terms of continuous arousal and valence values ranging between $-1$ and $1$. According to [20], the emotion values of consecutive frames do not change significantly. Hence, they split the movies into non-overlapping 5-seconds segments, in which the arousal and valence values are averaged over all frames in each segment. In this study, we use the same intended emotion annotations as in [20], [23] with one arousal and one valence value for each 5-second movie segment.

*b) MediaEval 2016 Emotional Impact of Movies Task (EIMT16):* The EIMT16 includes two subtasks namely global emotion prediction (Global EIMT16) and continuous emotion prediction (Continuous EIMT16). The purpose of the Global EIMT16 is to predict the valence and arousal for short movie excerpts, while the Continuous EIMT16 is created with the aim to predict those values continuously along the long video, which is similar to that in the extended COGNIMUSE dataset. This Global EIMT16 involves the largest number of movie excerpts in the LIRIS-ACCEDE database [2], and is complementary to COGNIMUSE dataset. Therefore, we choose the Global EIMT16 to conduct our experiments.

The Global EIMT16 includes $11,000$ movie excerpts, in which $9,800$ video excerpts are from the Discrete LIRIS-ACCEDE [17] and $1,200$ additional video excerpts [3]. The duration of each excerpt is between $8$ to $12$ seconds [17], which, as claimed by the authors, is long enough to induce emotion in the viewers and yet short enough to make the viewers feel only one emotion per excerpt. Each movie excerpt is annotated by only one score of arousal and valence values, which vary in the $[0, 5]$ range.

[2] https://liris-accede.ec-lyon.fr/

### B. Implementation details

*a) Feature extraction:* Movie excerpts in the Global EIMT16 have different duration and frame rates, hence, the FFMPEG tool is used to extract $64$ frames from each excerpt by setting the frame rate to $\lceil \frac{64}{t_i} \rceil$ (where $t_i$ is the duration of the $i$-th clip), which are then center-cropped to obtain a stack of $64$ frames with $224 \times 224$ pixels each. For the extended COGNIMUSE dataset, we use the same preprocessing techniques applied for the Global EIMT16 dataset. However, since the movies in the extended COGNIMUSE dataset has the same frame rate (25 fps), we use all $125$ frames in each 5-second excerpt instead of $64$ frames as in the Global EIMT16. Since the 5-second movie segments in the extended COGNIMUSE dataset are in chronological order, we use each movie segment as an element in the input sequence for the Temporal AttendAffectNet model. Movie excerpts in the Global EIMT16 are not in chronological order. Therefore, for the Temporal AttendAffectNet model, we split each movie excerpt, whose duration is between $8$ and $12$ seconds, into $4$ non-overlapping sub-segments with the same arousal and valence value for each. Each sub-segment is handled as a separate clip. Since the duration of each sub-segment is relatively short (between $2$ and $3$ seconds), we use the FFMPEG tool to extract only $16$ frames from each sub-segment, instead of $64$ for the whole excerpt.

*b) Optimization and training details::* We train the models to predict valence and arousal separately using the Adam optimizer with the loss function $L$ defined as follows:

$$L = MSE + (1 - \rho), \quad (3)$$

where $MSE$ and $\rho$ are the mean square error and Pearson correlation coefficient (PCC) respectively, which are computed from the predicted arousal/valence values and the ground truth.

We conduct experiments with a different number of heads, however, the use of a higher number of heads increases the computational cost and tends to cause overfitting during the model training. The models perform better with only two heads instead of eight heads as in the original Transformer model. In the feed-forward neural network sub-layers of the proposed models, we reduce the number of linear transformations from two to only one, which has the dimensionality of eight. For the extended COGNIMUSE dataset, the Feature AttendAffectNet is trained with a learning rate of $0.0005$, for a maximum of $500$ epochs, and a $0.1$ dropout ratio. The Temporal AttendAffectNet model is trained for a maximum of $1000$ epochs with a learning rate of $0.001$ and a $0.5$ dropout ratio. A batch size of $30$ and an early stopping criterion with a patience of $30$ epochs are set for both the models. For the Global EIMT16, the hyperparameters are the same as those for the extended COGNIMUSE dataset, except for the fact that the learning rate for both models is $0.01$, the batch size is $40$ and $20$ for arousal and valence prediction, respectively. The sequence length for the Temporal AttendAffectNet model is fixed to $4$ for the Global EIMT16 dataset and $5$ for the extended COGNIMUSE dataset. All of the models were implemented in Python 3.6 and the experiments were run on a NVIDIA GTX 1070.

TABLE I
ACCURACY OF THE PROPOSED MODELS ON THE EXTENDED COGNIMUSE DATASET.

| Models | Arousal | | Valence | |
| --- | --- | --- | --- | --- |
| | MSE | PCC | MSE | PCC |
| Feature AAN (only video) | 0.152 | 0.518 | 0.204 | 0.483 |
| Feature AAN (only audio) | 0.125 | 0.621 | 0.185 | 0.543 |
| **Feature AAN (video and audio)** | **0.124** | **0.630** | **0.178** | **0.572** |
| Temporal AAN (only video) | 0.178 | 0.457 | 0.267 | 0.232 |
| Temporal AAN (only audio) | 0.162 | 0.472 | 0.247 | 0.254 |
| **Temporal AAN (video and audio)** | **0.153** | **0.551** | **0.238** | **0.319** |
| Sivaprasad et al. [23] | **0.08** | **0.84** | **0.21** | **0.50** |

TABLE II
ACCURACY OF THE PROPOSED MODELS IN COMPARISON WITH STATE-OF-THE-ART ON THE GLOBAL EIMT16.

| Models | Arousal | | Valence | |
| --- | --- | --- | --- | --- |
| | MSE | PCC | MSE | PCC |
| Feature AAN (only video) | 0.933 | 0.350 | 0.764 | 0.342 |
| Feature ANN (only audio) | 1.111 | 0.397 | 0.209 | 0.327 |
| **Feature ANN (video and audio)** | **0.742** | **0.503** | **0.185** | **0.467** |
| Temporal ANN (only video) | 1.182 | 0.151 | 0.256 | 0.190 |
| Temporal ANN (only audio) | 1.159 | 0.185 | 0.225 | 0.285 |
| **Temporal ANN (video and audio)** | **0.854** | **0.210** | **0.218** | **0.415** |
| Liu et al. [56] | 1.182 | 0.212 | 0.236 | 0.379 |
| Chen et al. [55] | 1.479 | 0.467 | 0.201 | 0.419 |
| Yi et al. [22] | 1.173 | 0.446 | 0.198 | 0.399 |
| Yi et al. [41] | **0.542** | **0.522** | **0.193** | **0.468** |

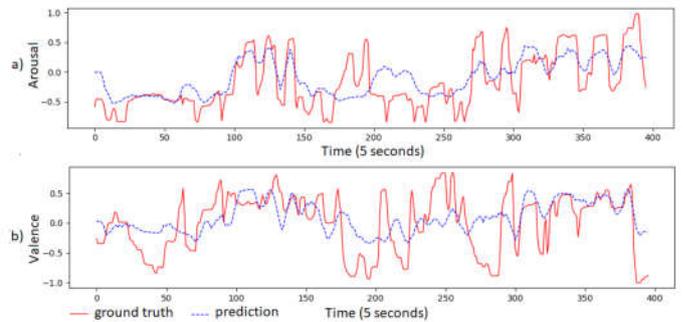

Fig. 5. Arousal (a) and valence (b) values for the "Million Dollar Baby" movie clip.

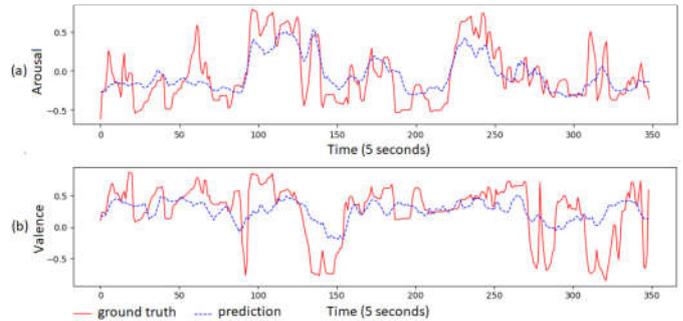

Fig. 6. Arousal (a) and valence (b) values for the "Ratatouille" movie clip.

In the training stage of the Temporal AttendAffectNet, the available previous outputs are used as the decoder input, and the model is parallelized. In the evaluation stage, outputs are not given, therefore, we re-feed our model for each position of the output sequence until we come across the "end-of-sequence" (due to the space constraints, we refer to the Appendix for more details on the evaluation stage of the Temporal AttendAffectNet).

### C. Evaluation metrics

The proposed models are evaluated in terms of the MSE and PCC between the predicted arousal/valence values and the ground truth. The extended COGNIMUSE dataset includes twelve movie clips, therefore, we conduct leave-one-out cross-validation, then the MSE and PCC are computed for all predicted values as used in [23].

### V. EXPERIMENTAL RESULTS

We trained and validated the proposed models on the extended COGNIMUSE and the Global EIMT16 datasets. The results are summarised in Table I and Table II, respectively.

Based on the MSE and PCC, we observe that the Feature AttendAffectNet has a higher accuracy for predicting emotions of movie viewers than the Temporal AttendAffectNet. This might be explained by the fact that 5-second movie segments in the extended COGNIMUSE dataset and movie excerpts in the Global EIMT2016 are long enough to induce a specific emotion; and the feature sets extracted from these movie excerpts could be sufficient to carry the emotional content.

The proposed models based on the audio stream only have a higher prediction accuracy than those based on the video stream only. This phenomenon was also observed in [24]. This suggests that audio has a larger influence on evoked emotion than video in multimedia content. The clips in the extended COGNIMUSE and Global EIMT16 datasets are extracted from movies, therefore, they include sound effects, speech and music. Filmmakers purposely select audio to convey the emotional meaning of a story or to deliver clear semantic messages to movie viewers. Music plays an important role in this, as it has long been known for its ability to influence emotions [47]. Another reason could be because the extracted audio features are simply more representative of emotion than the extracted video features. The models that use both the visual and audio feature sets obtain the highest accuracy for both arousal and valence prediction, and the self-attention mechanism is able to combine them in an effective manner.

*a) Comparison to state-of-the-art results:* The Feature AttendAffectNet for valence prediction outperforms the LSTM approach from [23] on the extended COGNIMUSE. Note that in [23], in addition to applying correlation-based feature selection [54] to select audio and video features, late fusion is also used to obtain the estimated arousal and valence values. We tried applying late fusion in our approach, our results were not significantly higher, at a much higher training cost.

We also compare AttendAffectNet to state-of-the-art models on the Global EIMT16. Using both video and audio, the Feature AttendAffectNet outperforms top 3 results [22], [55],

[56] in both arousal and valence prediction. The MSE and PCC are 0.742 and 0.503 respectively for arousal; 0.185 and 0.467 respectively for valence. These MSE and PCC are almost equivalent to those in [41], except for the MSE for arousal. Note that in this study, we only use the element-wise averaging, while Yi et al. [41] use the concatenation of mean and standard deviation of the extracted feature vectors to represent for each modality.

*b) Visualization of predicted arousal and valence:* We visualize the predicted values of arousal and valence using the Feature AttendAffectNet model and their ground truth for a usual movie clip named "Million Dollar Baby" and an animated one called "Ratatouille" in Figures 5 and 6, respectively. The predicted values of arousal follow the ground truth well for both movies. For "Million Dollar Baby", the PCCs for arousal and valence are 0.715 and 0.507 respectively, and MSEs are 0.111 and 0.184 respectively. For "Ratatouille", 0.774 and 0.646 are the PCCs for arousal and valence respectively, and the corresponding MSEs are 0.054 and 0.172.

## VI. Conclusion

This study presents AttendAffectNet, a multimodal approach based on the self-attention mechanism, which allows to find unique and cross-correspondence contributions of features extracted from multiple modalities to predict emotions of viewers from movies.

We use pre-trained deep neural networks and the OpenSMILE toolkit to extract a wide range of features from audio and video streams. We compare different ways to integrate them using the self-attention based network. First, we look at the use of the self-attention mechanism to attend to different features, and secondly, we use it to look back over time at the movie. We analyse the performance of our proposed AttendAffectNet model by training it with self-attention for feature and temporal components. Our results on the extended COGNIMUSE and Global EIMT16 datasets show that the former performs better than the latter. This could be due to the fact that movie segments are already long enough to convey emotional content.

In addition, we notice that the AttendAffectNet trained on audio features outperforms that on video features. This might be due to either the stronger influence of audio on emotion than video, or the fact that our audio features are better able to represent the emotions. Both models reach the highest performance when all audio and video features are combined.

### Acknowledgement

This work is supported by MOE Tier 2 grant no. MOE2018-T2-2-161 and SRG ISTD 2017 129.